\documentclass[prb,showpacs,twocolumn,floatfix,amsmath,amsfonts]{revtex4}
\usepackage{graphicx}
\usepackage{epstopdf}
\usepackage{textcomp}
\usepackage[utf8]{inputenc}
\usepackage[russian,english]{babel}
\usepackage[usenames, dvipsnames]{color}
\usepackage{hyperref}
\usepackage{color}

\begin{document}

\title{Energy transfer in strained graphene assisted by discrete breathers \\ excited by external ac driving}

\author{Iman~Evazzade$^1$}
\email{i.evazzade@mail.um.ac.ir}

\author{Ivan~P.~Lobzenko$^2$}
\email{ivanlobzenko@gmail.com}

\author{Elena~A.~Korznikova$^2$}
\email{elena.a.korznikova@gmail.com}

\author{Ilya~A.~Ovid'ko$^3$}
\email{ovidko@gmail.com}

\author{Mahmood~Rezaee~Roknabadi$^1$}
\email{roknabad@um.ac.ir}

\author{Sergey V. Dmitriev$^{2,3}$}
\email{dmitriev.sergey.v@gmail.com}

\affiliation{$^1$Department of Physics, Faculty of Science, Ferdowsi University of Mashhad, Mashhad, Iran \\
$^2$Institute for Metals Superplasticity Problems, Russian Academy of Sciences, Ufa, 450001 Russia \\
$^3$Research Laboratory for Mechanics of New Nanomaterials, Peter the Great
St. Petersburg Polytechnical University, St. Petersburg 195251, Russia}

\date{\today}

\begin{abstract}
In the present molecular dynamics study, external ac driving is used at frequencies outside the phonon spectrum to excite gap DBs in uniformly strained graphene nanoribbon. Harmonic displacement or harmonic force is applied to a zigzag atomic chain of graphene. In the former case non-propagating DBs are excited on the atoms next to the driven atoms, while in the latter case the excited DBs propagate along the nanoribbon. The energy transfer along the nanoribbon assisted by the DBs is investigated in detail and the differences between harmonic displacement driving and harmonic force driving are discussed. It is concluded that the amplitude of external driving at out of phonon spectrum frequencies should not necessarily be large to obtain a noticeable energy transfer to the system. Overall, our results suggest that external harmonic driving even at relatively small driving amplitudes can be used to control excitation of DBs and consequently the energy transfer to the system.
\end{abstract}


\maketitle

\section{Introduction}

Discrete breather (DB), also called intrinsic localized mode, is a spatially localized, time-periodic excitation in a defect-free discrete nonlinear lattice. The concept of DB has emerged in nonlinear science as a mathematical discovery three decades ago~\cite{Dolgov1986,ST1988,Page1990}. Later DBs were directly observed in many macroscopic and mesoscopic systems of different physical nature ~\cite{Morandotti1999,Binder2000,kivsharRev2004,SatoRMP2006,Flach2008}. They have also been reported by experimentalists in different crystals, which can be regarded as nonlinear lattices if considered at the atomic scale~\cite{dbPtCl1999,dbPtCl2002,manleyU2006,manleyU2008,manleyNaI2009,Archilla2015,DmitrievUFN2016}.

Concentration of large-energy DBs in crystals in thermal equilibrium increases with temperature \cite{Ivanchenko2004,Eleftheriou2005,KhaDmi2D2011}, but it remains rather small for unambiguous experimental detection \cite{siversNaI2013}. However in non-equilibrium states DB concentration can be orders of magnitude higher than in equilibrium state. For example, DBs can emerge spontaneously as a result of the modulational instability of a short-wavelength extended vibrational mode \cite{Dauxois2005,DNY2009,DSPKIK2009,dmitrievMDNaCl2010}. They can also be excited by external driving at frequencies outside the phonon spectrum and sufficiently large driving amplitude, according to the so-called supratransmission phenomenon \cite{GL2002,KLR2004,KTDBA2004}. In the theoretical work by R\"{o}ssler and Page the possibility of optical creation of DBs in crystals has been demonstrated \cite{RP2000}. The authors have shown that DBs can be excited directly by applying a sequence of femtosecond visible laser pulses at THz repetition rates or indirectly via decay of an unstable extended lattice mode optically excited by a single picosecond far-infrared laser pulse with linearly chirped frequency. Recent advances in laser pulse shaping make these approaches experimentally promising. Thus, it is interesting to study basic mechanisms of DB excitation in crystals by external driving.

Recently two-dimensional crystals such as graphene \cite{Graphene}, graphane \cite{Graphane}, MoS$_2$ \cite{MoS2}, phosphorene \cite{Phosphorene}, silicene \cite{Silicene} and others have attracted enormous attention of researchers because they have unique combination of physical and mechanical properties promising for a number of applications. It has been shown that graphene and graphane support DBs \cite{RAMS,LiuBo,Chechin2014,Baimova2015,BaimovaGR}. The first theoretical studies on DBs in graphene and carbon nanotubes have been done by Japanese researchers \cite{Japan1,Japan2,Japan3,Japan4,Japan5,Japan6}. In their studies the DBs with frequencies above the phonon spectrum were identified. Such DBs in unstrained graphene and carbon nanotubes are linearly unstable \cite{Japan5,Japan6}. It is well-known that graphene does not possess a gap in the phonon spectrum, but a gap can be induced by application of homogeneous elastic strain and then robust gap DBs (having frequencies within the phonon gap) can be excited \cite{KhaDmiKiv2011,Korznikova1,Korznikova2,myAbGrE2016}. DB in unstrained graphene with out-of-plane oscillations has been recently reported by Hizhnyakov {\it et al.} \cite{DBnormal}. The DB demonstrates hard type nonlinearity (its frequency increases with the amplitude) with frequencies above the out-of-plane phonon spectrum but within the in-plane phonon spectrum. For the unstrained graphene the localized vibrational mode with in-plane atomic vibrations has been recently reported based on molecular dynamics simulations with the Tersoff potential \cite{Tsir}. The mode has frequencies above the phonon spectrum. However, the reported mode is not actually a DB but rather it is a defect mode. The Tersoff potential supports a stable defect in graphene with single valence bond longer than the other bonds and this long bond can vibrate with the above spectrum frequencies. That is why in the present molecular dynamics study based on the Savin potential \cite{savinCHPot2010} we focus on the gap DBs in strained graphene. Importantly, the existence of the gap DBs with in-plane vibrations in strained graphene has been confirmed with the aid of {\it ab initio} simulations based on the DFT theory \cite{myAbGrE2016}.

A number of experimental and especially theoretical works have shown that DBs do exist in various crystals \cite{dbPtCl1999,dbPtCl2002,manleyU2006,manleyU2008,manleyNaI2009,DmitrievUFN2016,kiselevMD1997,voulgarakisSi2004,hizhnyakovNiNb2011,Murzaev2016,Medvedev2015}. It is very timely to explain the role of DBs in the formation of the properties of real crystals and to possibly develop applications of DBs in new technologies. There exist pioneering works addressing these ambitious objectives. Velarde with co-authors argue that the localized excitations can contribute to the transport of electric charge \cite{Velarde1,Velarde2,Velarde3}. DBs excited during low energy plasma surface treatment can participate in annealing of defects in single crystal germanium improving its quality for applications \cite{Archilla2015}.

DBs are nonlinear vibrational modes and thus, they are most likely excited when the crystal receives energy in large portions, e.g., during optical driving, irradiation, under a high-density electric current, plastic deformation, etc. \cite{RP2000,Archilla2015,Du1,Du2}. In the physical experiments, in the presence of inevitable perturbations, it is only possible to have quasi-breathers with finite lifetime and not exact time periodicity \cite{chechinPRE2006,chechinNLN2009}.

Thermal conductivity of two-dimensional nanomaterials is interesting from both theoretical and practical standpoints. Such systems may violate the Fourier's law of thermal conductivity (see Refs. \cite{savinCHPot2010,DXiong2014} and references therein). They allow manipulation of thermal conductivity by means of doping \cite{Bliu2014,YGuo2016}, introduction of defects and elastic strain engineering \cite{ZDing2015}, through design of various heterostructures \cite{Bliu2012,YZhang2016,BLiuGreSil2014}. Possible effect of DBs on thermal conductivity has been discussed in a recent theoretical work \cite{DXiong2016}. 

In our preliminary study it was shown that DBs can affect the energy transfer in graphene under ac driving~\cite{myLOMgreTransf2016}. We present here a detailed discussion on that problem.

The present paper is devoted to the investigation of energy transfer mechanisms in graphene nanoribbons under external driving  by means of classical molecular dynamics. The paper is organized as follows. The computational model and the simulation setup are described in Sec.~\ref{SimulationSetup}. Energy transfer along the graphene nanoribbon under displacement driving and force driving is analyzed in Sec.~\ref{DisplDriving} and Sec.~\ref{ForceDriving}, respectively. The results are discussed and concluded in Sec.~\ref{Conclusion}.

\begin{figure}
\includegraphics*[scale=1.7]{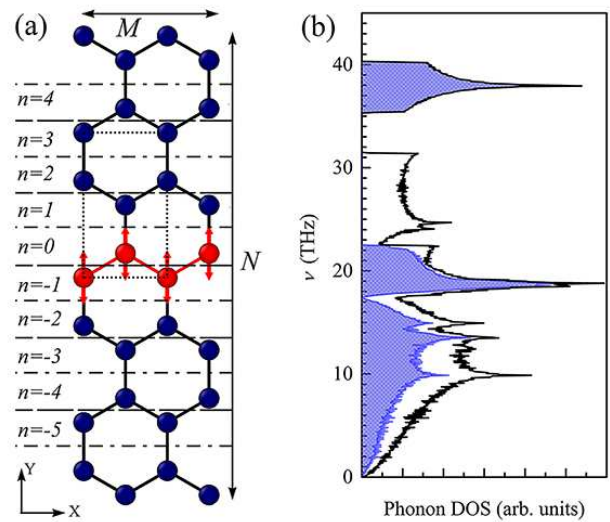}
\caption{(Color online) (a) Crystal lattice of graphene. Dotted line shows the rectangular translational cell containing four carbon atoms. Horizontal atomic rows are numbered with the index $n$.
(b) Phonon density of states for the system under the elastic strain with the components $\varepsilon_{xx} = 0.35 $, $\varepsilon_{yy} = -0.1 $, and $\varepsilon_{xy} = 0$.} \label{glat}
\end{figure}

\section{Simulation setup}
\label{SimulationSetup}

A graphene sheet [see Fig.~\ref{glat}(a)] is a two-dimensional hexagonal crystal with a primitive translational cell containing two carbon atoms. We consider rectangular translational cell with four atoms as shown by the dotted line. To simulate the nanoribbons, a rectangular supercell with dimensions $M\times N$ was built, where $M$ and $N$ are numbers of translational unit cells along $x$ (zigzag graphene direction) and $y$ (armchair graphene direction) axes, respectively. Horizontal atomic rows are numbered with the index $n$. The periodic boundary conditions are applied to the simulation supercell.

Interatomic interactions are described by the set of interatomic potentials developed in \cite{savinCHPot2010}. The equilibrium valent bond length in the unstrained graphene is $\rho_0 = 1.418$~\AA. A distinctive feature of the potentials is that they reproduce the dispersion curve of graphene better than the Brenner potentials~\cite{brennerPot1990}.

In order to induce a gap into phonon spectrum, we apply in-plane elastic strain with the components $\varepsilon_{xx} = 0.35 $, $\varepsilon_{yy} = -0.1 $, and $\varepsilon_{xy} = 0$. Most of the results are obtained for these values of strain but we also report on the effect of the elastic strain on the studied phenomena. The equilibrium positions of atoms in uniformly strained graphene are found by minimizing the potential energy of the crystal. For the chosen strain components the equilibrium flat configuration of graphene is stable~\cite{baimova2012}. It should be pointed out that the maximal level of strain used in our simulations is very high and it is at the stability border of graphene reported in the literature. For example, quantum mechanics and quantum molecular dynamics calculations \cite{Gao2009} gave the limiting values for uniaxial strain along zigzag (armchair) direction $\varepsilon_{xx} = 0.38$ ($\varepsilon_{yy} = 0.19$). Density functional perturbation theory was employed to calculate the dispersion curves of uniaxially loaded graphene and the phonon instability was found at $\varepsilon_{xx} = 0.266$ for the zigzag direction and $\varepsilon_{yy} = 0.194$ for the armchair direction \cite{Liu2007}. In molecular dynamics study for graphene nanoribbons oriented along the armchair direction the critical strain of $\varepsilon_{yy} = 0.30$ was reported \cite{Bu2009}. The aim of using such a high maximal elastic strain used in our simulations is to enhance the reported effects. As it will be demonstrated, similar effects are observed at much smaller values of strain but with a reduced clarity. 

The phonon density of states (DOS) of our system under the strain is depicted in Fig.~\ref{glat}(b). It is separated into the DOS for in-plane modes (hatched) and out-of-plane modes (unhatched). A wide gap in $xy$ phonon DOS can be seen, which makes it possible for in-plane DBs to exist with frequencies inside the gap (see~\cite{KhaDmiKiv2011} for the thorough discussion).

To start an energy transfer to the system we introduce a harmonic external driving of the atoms belonging to one zigzag chain of the carbon atoms in the middle of the ribbon. These atoms belong to the atomic rows with the numbers $n=-1$ and $n=0$ [see Fig.~\ref{glat}(a)]. The number of atoms in that chain equals $2M$. We choose two most simple types of the driving, namely, displacement driving and force driving. For the former case the prescribed displacements of the atoms change in time according to
\begin{equation}
\label{eq1}
\Delta y_0(t)=\Delta y_{-1}(t)=A{\sin}(2\pi{\nu}t),
\end{equation}
where ${\Delta}y$ represents the displacement of atoms from their equilibrium positions in $y$ direction, $A$ is the amplitude in Angstrom and $\nu$ is the frequency in THz. In the case of force driving, to each atom of the driven zigzag chain the external force is applied as follows
\begin{equation}
\label{eq2}
Y_0(t)=Y_{-1}(t)=F{\sin}(2\pi{\nu}t),
\end{equation}
here $Y$ indicates the external force in $y$ direction and $F$ is the amplitude of the force in nN.

At $t=0$ all atoms are in their equilibrium positions with zero initial velocities.

Each carbon atom in our simulations has three degrees of freedom, but due to the symmetry of external driving, the atoms predominantly move along $y$ axis. At very large amplitudes of the external driving the motion of atoms along $y$ axis can become unstable. We limit the values of the driving amplitudes and the simulation time in order to study the quasi-one-dimensional problem with atoms moving along $y$ axis with negligible components of displacements along $x$ and $z$ axes.

To check the effect of the computational cell size effect the results were compared for $M\times N=1\times 200$, $1\times 400$, and $4\times 400$. It was found that the results for $M=1$ and $M=4$ coincide and the results for $N=200$ and $N=400$ are very close. Most of the results reported in this study were obtained for $M\times N=1\times 400$.

The simulation time is equal to the time needed for any perturbation from the energy source to reach the ends of the computational cell. This limitation ensures the absence of the effect of interference of the waves travelling from the source in the opposite directions.

The quantity which describes the energy transfer to the system is power per one translational cell in $x$ direction. Power as the function of time can be calculated as
\begin{equation}
\label{eq3}
P(t)=\frac{1}{M}\frac{dE(t)}{dt},
\end{equation}
where $E$ is the total (kinetic plus potential) energy of the system. In the present work we average the power given by Eq.~(\ref{eq3}) over the entire simulation time.

\section{Displacement driving}
\label{DisplDriving}

\begin{figure}
\includegraphics*[scale=0.35]{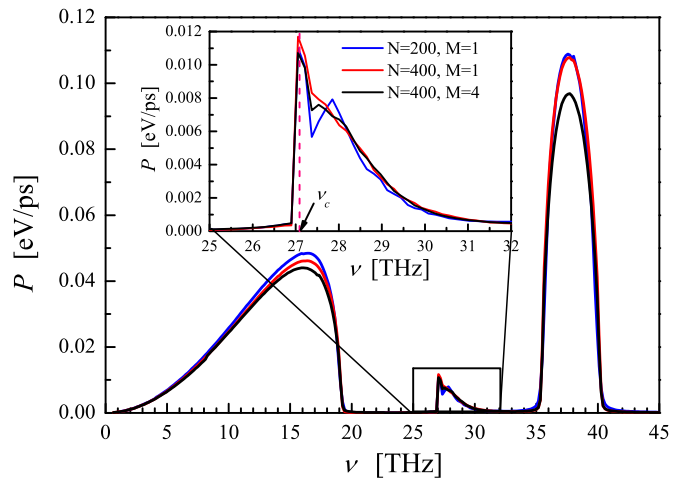}
\caption{\label{gPDisp}(Color online) The dependence of power $P$ averaged over entire simulation time on driving frequency $\nu$ with $A=0.01$~\AA$\,$ for $M=1$, $N=200$ (blue line), $M=1$, $N=400$ (red line), $M=4$, $N=400$ (black line). Dashed line shows the critical frequency $\nu_c=27.1$~THz.}
\end{figure}

For the displacement driving we apply Eq.~(\ref{eq1}) with different frequencies $\nu$ and amplitudes $A$ of the driving. In Fig.~\ref{gPDisp} the dependence of the power $P$ on driving frequency $\nu$ is shown for the amplitude $A=0.01$~\AA. To demonstrate the computational cell size effect the results are compared for $N=200$, $M=1$ (blue line); $N=400$, $M=1$ (red line); and $N=400$ and $M=4$ (black line). The larges difference between the blue and red curves within the phonon spectrum is observed at about 16~THz and it does not exceed 5\%. The maximal difference between these curves within the gap is about 25\% and it is due to the higher random oscillation of the curve for $N=200$. For $N=400$ the curve is smooth and that is why this length of the nanoribbon was chosen for simulations. As for the black curve, within the gap it is close to the red curve. The largest difference between the black and red curves is observed in the optic phonon band at 37.5~THz (about 12\%). This difference is attributed to the excitation of displacements of atoms in $x$-direction in the case of $M=4$, which are suppressed in the case of $M=1$. As evident from Fig.~\ref{gPDisp}, such displacements are mostly excited at high frequencies and they do not affect the most interesting for this study frequency range within the gap of the phonon spectrum. In what follows the results for $N=400$ and $M=1$ are reported.

The dependence of power on frequency shown in Fig.~\ref{gPDisp} should be compared with the phonon DOS (Fig.~\ref{glat}b). The phonon DOS for in-plane modes has frequencies in the ranges $0\le \nu \le 22.5$~THz and $35.4\le \nu \le 40.3$~THz. As expected, the energy transfer to the system takes place within these regions by means of excitation of the travelling small-amplitude phonon waves. Besides there is the frequency region where energy transfer to the system takes place ($P>0$) but the driving frequency does not match with any in-plane phonon frequency (in the gap of DOS for in-plane phonon modes). The inset in Fig.~\ref{gPDisp} shows the magnification of the frequency region with $P>0$ in the gap of in-plane phonon DOS. It is interesting to reveal the mechanism of energy transfer to the system within this frequency range. For this we analyse the $y$-displacements of particles as the functions of time.

Firstly, referring to the inset of Fig.~\ref{gPDisp}, we note the existence of the critical driving frequency $\nu_c=27.1$~THz, such that for $\nu<\nu_c$ power $P$ is positive but relatively small and it sharply increases at $\nu=\nu_c$, going down for $\nu>\nu_c$.
\begin{figure}[!h]
\includegraphics*[scale=0.32]{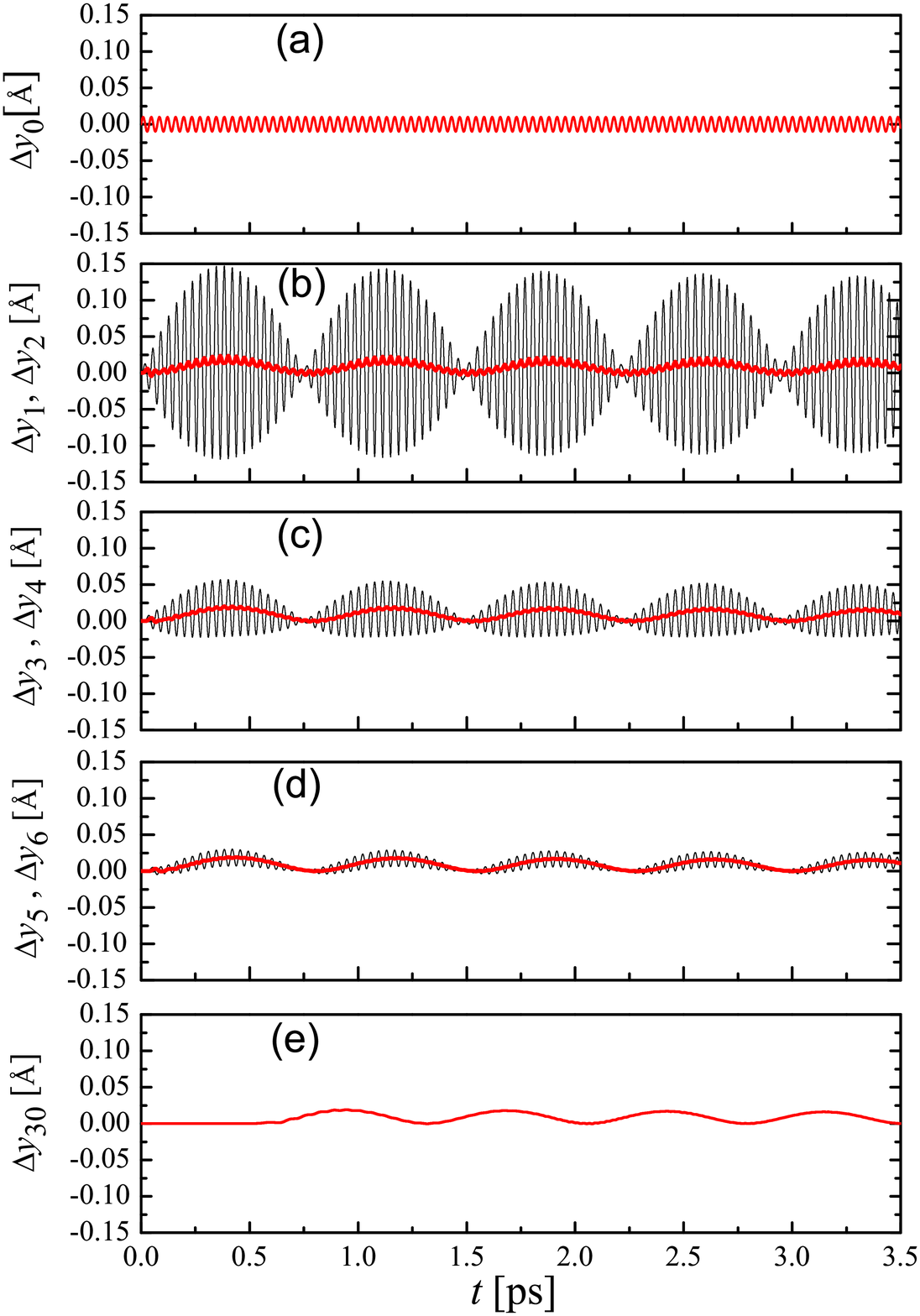}
\caption{\label{g-osc1}(Color online) Displacements in $y$ direction as the functions of time for (a) driven atom, $n=0$, (b) atoms $n=1$ and 2, (c) atoms $n=3$ and 4, (d) atoms $n=5$ and 6, (e) atom $n=30$. Displacement driving amplitude is $A=0.01$~\AA~and frequency is $\nu=26.7$~THz, which is slightly below the critical frequency of $\nu_c=27.1$~THz. Curves for odd (even) atoms are shown in black (red).}
\end{figure}
In Fig.~\ref{g-osc1} the time evolution of $y$-displacements is shown for the driven atom ($n=0$), its six neighbors with $n=1,...,6$, and for the remote atom with $n=30$. This result corresponds to the driving amplitude $A=0.01$~\AA$\,$ and frequency $\nu=26.7$~THz, which is slightly below $\nu_c$. As shown in (b), the $n=1$ particle oscillates with the amplitude modulated in time with the maximal value one order of magnitude greater than the amplitude of the driven atom. This excited atom can be regarded as the large-amplitude DB whose amplitude (quasi)periodically changes in time. Atoms with odd $n$ have vibration amplitudes larger than the nearest atoms with even $n$. This is understandable taking into account that in graphene [see Fig.~\ref{glat}(a)] for even $n$ the valence bonds connecting atoms $n$ and $(n+1)$ are oriented along $y$ axis, while for odd $n$ they are tilted. This means that the stiffness of the bonds in vertical direction alternates between the large stiffness of a vertical bond and a smaller stiffness of a pair of tilted bonds. Stiff bonds can transmit high-frequency vibrations, while soft bonds transmit the low-frequency envelop vibrations that arise from the anharmonicity of the DB with the modulated amplitude. As a result, the energy flows from the driven atoms into the crystal by means of the low-frequency running phonon waves with the period equal to the period of modulation of the DB amplitude. This is clearly seen in Fig.~\ref{g-osc1}(e) where displacement of the remote particle is shown.

\begin{figure}
\includegraphics*[scale=0.35]{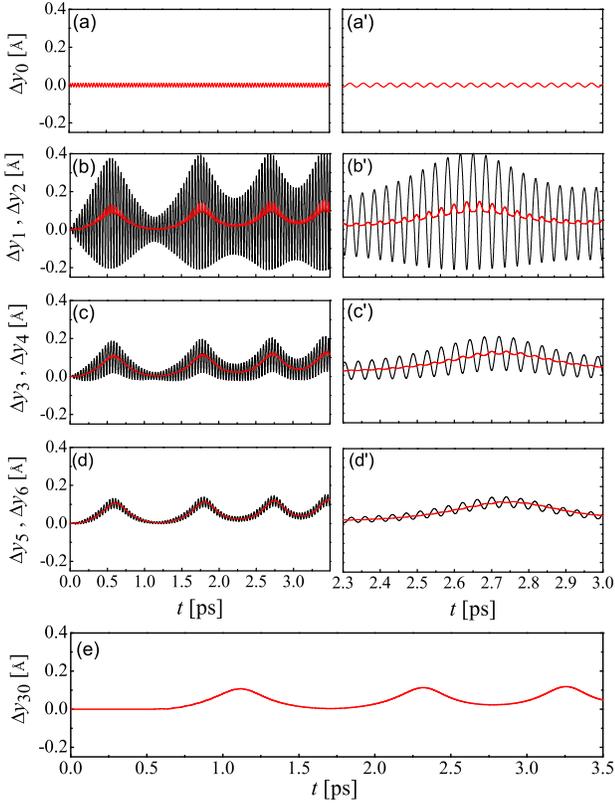}
\caption{\label{g-osc2}(Color online) Same as in Fig.~\ref{g-osc1} but for the driving frequency of $\nu=27.7$~THz, which is slightly above $\nu_c=27.1$~THz. Left column shows the entire simulation time domain, while the right column only its final part.}
\end{figure}

\begin{figure}
\includegraphics*[scale=0.35]{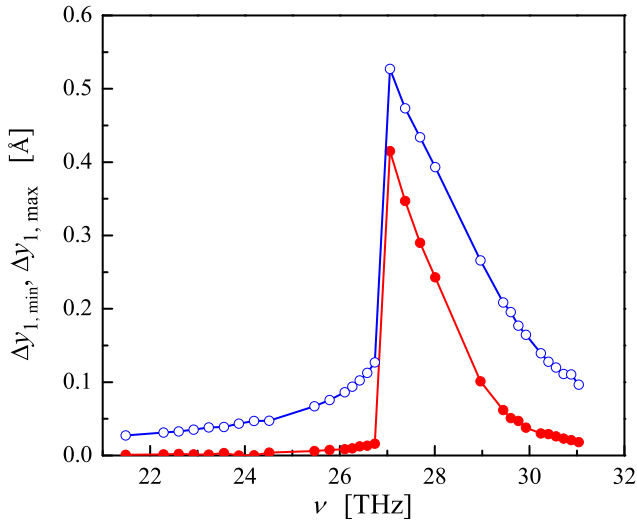}
\caption{\label{g-ampD}(Color online) Maximum (open dots) and minimum (dots) amplitude of oscillations of the atoms in the row $n=1$ versus driving frequency. Displacement driving amplitude is $A=0.01$~\AA.}
\end{figure}

The behaviour of the system changes dramatically when the driving frequency passes the value of $26.9$~THz. In Fig.~\ref{g-osc2} the same as in Fig.~\ref{g-osc1} is shown for the driving frequency of $\nu=27.7$~THz and the same driving amplitude of $A=0.01$~\AA. The left column shows the entire simulation time domain, while the right column presents the final part of the simulation time domain in an enlarged scale. Note that the maximal oscillation amplitude of atom $n=1$ in Fig.~\ref{g-osc2}(b) is about three times larger than in Fig.~\ref{g-osc1}(b). It is also clear that the minimal amplitude of $n=1$ atom in Fig.~\ref{g-osc2}(b) does not drop to zero as it was for the case shown in Fig.~\ref{g-osc1}(b). In Fig.~\ref{g-ampD} the maximal (open dots) and minimal (dots) values of the oscillation amplitude for $n=1$ atom are shown. Below (above) $\nu_c$ the minimal value is nearly zero (is non-zero) and the maximal value is relatively small (large).

\begin{figure}
\includegraphics*[scale=0.35]{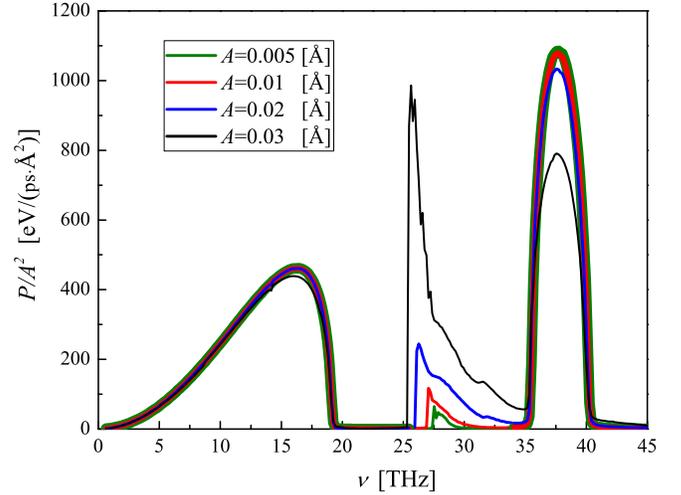}
\caption{\label{g-PDisp2}(Color online) Ratio $P/A^2$ as the function of driving frequency $\nu$ for the driving amplitudes $A=0.005$~\AA$\,$ (green line), $A=0.01$~\AA$\,$ (red line), and $A=0.02$~\AA$\,$ (blue line).}
\end{figure}

We have also checked how the energy source power $P$ depends on the driving amplitude $A$. Since phonon energy is proportional to $A^2$, it is expected that the ratio $P/A^2$ should be constant for given driving frequency $\nu$. In Fig.~\ref{g-PDisp2} the ratio $P/A^2$ is presented as a function of $\nu$ for $A=0.005$~\AA$\,$ (green line), $A=0.01$~\AA$\,$ (red line), $A=0.02$~\AA$\,$ (blue line), and $A=0.03$~\AA$\,$ (black line). Indeed the lines overlap within both in-plane phonon bands for the small amplitude phonons, but with increasing phonon amplitude $P/A^2$ starts to reduce, especially for the optic phonon band. Note that within the gap of in-plane phonon DOS the ratio $P/A^2$ increases with increasing $A$ even for small driving amplitudes. At the same time, the critical frequency $\nu_c$ of sharp increase in power shifts to the smaller values for larger $A$. This result suggests that for larger driving amplitudes the contribution of DB to the energy transport increases in comparison to phonons.

The appearance of the critical frequency $\nu_c$ within the gap of the in-plane phonon DOS can be explained by the calculation of the DB frequency for different driving frequencies $\nu$. It turns out that the DB frequency practically {\em does not depend} on $\nu$ for given driving amplitude $A$. DB frequency also weakly depends on time in spite of the fact that its amplitude varies with time [see Fig.~\ref{g-osc2}(b)].

\begin{figure}[!t]
\includegraphics*[scale=0.35]{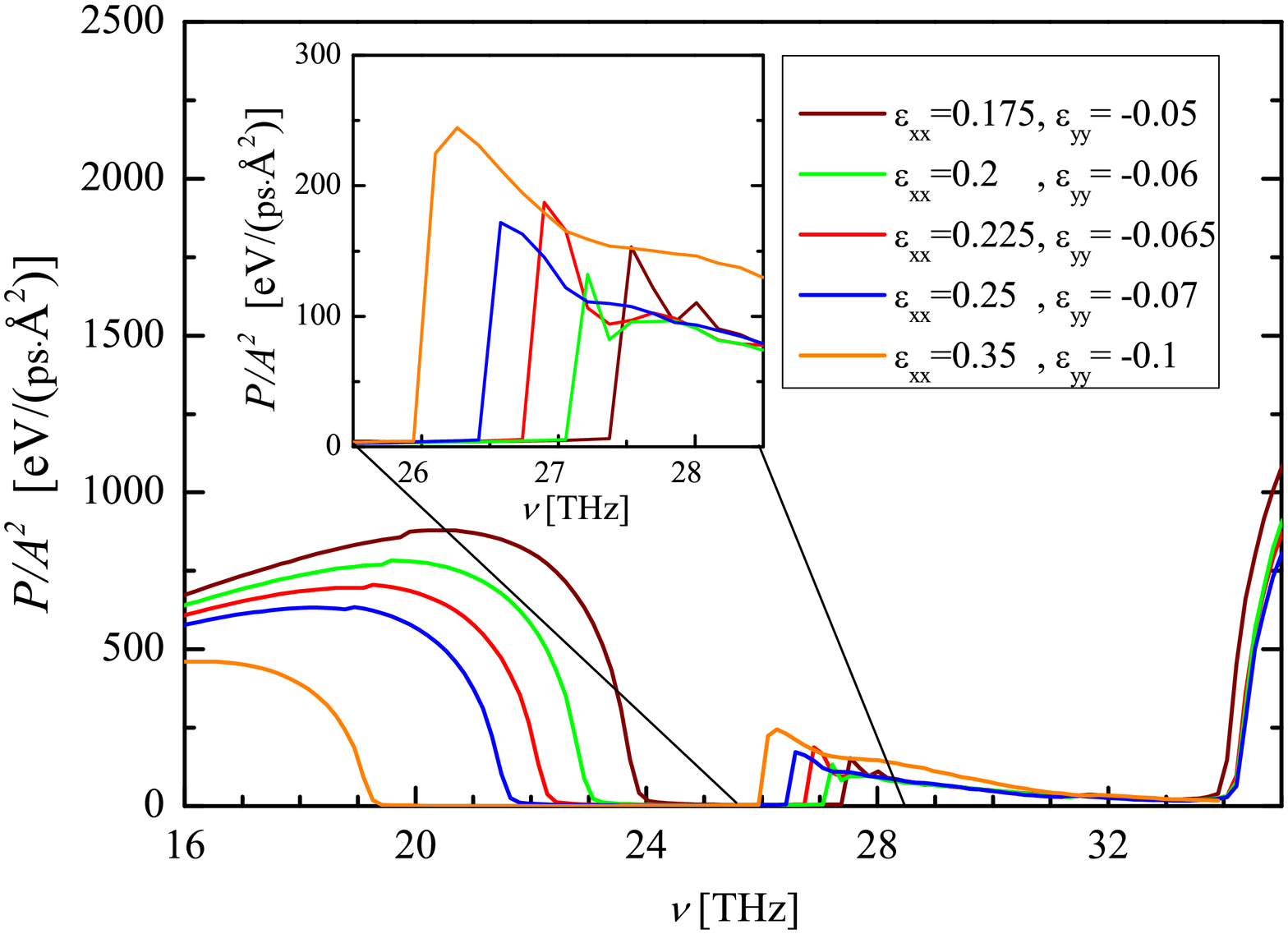}
\caption{\label{g-strain}(Color online) The dependence of power $P$ averaged over entire simulation time on driving frequency $\nu$ for different strain of the ribbon. Driving amplitude is $A=0.01$~\AA.}
\end{figure}

In the Table~\ref{t-nu} one can see the critical frequencies and corresponding power rates for different driving amplitudes. Red shift of $\nu_c$ with increasing driving amplitude $A$ is expected taking into account soft type of nonlinearity of the DB in strained graphene (with increasing amplitude, DB frequency decreases) \cite{KhaDmiKiv2011}. We need to point that all above results were obtained for graphene nanoribbons under the elastic strain with the components $\varepsilon_{xx}~=~0.35 $, $\varepsilon_{yy}~=~-0.1$. In purpose to show how the critical frequency depends on the strain parameters, we plot Fig.~\ref{g-strain}. It is clear that with decrease of strain components $\nu_c$ increases. This is related to the asymmetric effect of the elastic strain on the edges of the phonon spectrum gap. Decreasing strain results in a blue shift of the lower edge of the gap with very small effect on the upper edge. As a result, frequency of the center of the gap increases with decreasing strain and so does the critical driving frequency, which lies in the middle part of the gap. 

For small frequencies ($\nu<5$~THz) one has quadratic dependence of power on the driving frequency, see Fig.~\ref{g-PDisp2}. This is understandable taking into account that the low-frequency (sound) phonons are dispersionless and their energy density is proportional to $A^2\nu^2$. Displacement driving with the amplitude $A$ produces phonons with the same amplitude regardless the driving frequency $\nu$. For constant driving amplitude the power is thus proportional to $\nu^2$. For driving frequencies higher than 5~THz phonon speed decreases due to the dispersion effect and the power increases with $\nu$ slower than $\nu^2$. When the driving frequency approaches the edge of the acoustic band frequency, the power drops down to zero.

\begin{table}[!h]
\caption{\label{t-nu}Critical values of frequency for different driving amplitudes}
\begin{center}
\begin{tabular}{ l  c  c }
\hline
Amplitude [\AA]    & $\nu_c$ [THz]     & Power [eV/(ps${\cdot}$\AA$^2$)] \\
\hline
\hline
$0.03$   & $25.6$ & $985.06$ \\
$0.02$   & $25.6$ & $241.77$\\
$0.01$   & $27.1$ & $115.11$\\
$0.005$  & $27.6$ & $66.89$\\
\hline
\end{tabular}
\end{center}
\end{table}

\begin{table}[!h]
\caption{\label{table}Critical values of frequency for different elastic strains}
\begin{center}
\begin{tabular}{ l  c  c }
\hline
$\varepsilon_{xx}$  & $\nu_c$ [THz]     & Power [eV/(ps${\cdot}$\AA$^2$)] \\
\hline
\hline
$0.175$   & $27.53$ & $157.19$ \\
$0.2$   & $27.22$ & $131.87$\\
$0.225$   & $26.9$ & $186.58$\\
$0.25$ & $26.58$ & $170.68$\\
$0.275$   & $26.42$ & $1713.35$\\
$0.3$   & $26.41$ & $1603.93$\\
$0.325$ & $26.26$ & $608.37$\\
\hline
\end{tabular}
\end{center}
\end{table}

\section{Force driving}
\label{ForceDriving}

\begin{figure}
\includegraphics*[scale=0.32]{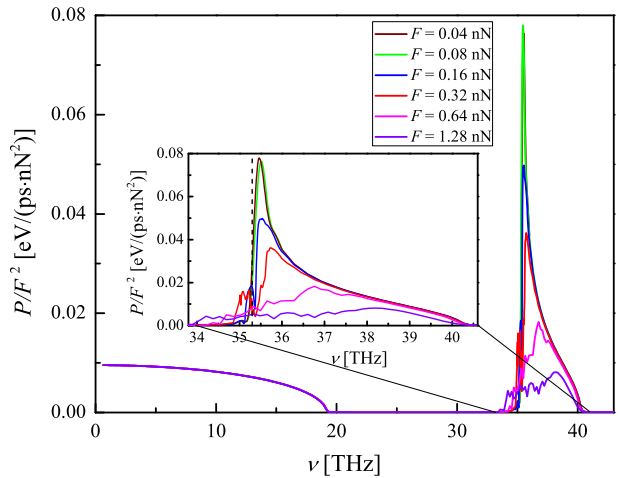}
\caption{\label{g-PForce}(Color online) The dependence of the normalized power $P/F^2$ on driving frequency $\nu$ in the case of force driving for the amplitudes $F=0.04~nN$ (brown), $F=0.08~nN$ (green), $F=0.16~nN$ (blue), $F=0.32~nN$ (red), $F=0.64~nN$ (magenta), and $F=1.28~nN$~(violet). The greater is $F$ the lower is the curve. The inset shows the magnification of the optical frequency band. Vertical dashed line in the inset indicates the upper edge of the phonon gap at 35.37~THz.}
\end{figure}

For the case of force driving Eq.~(\ref{eq2}) the power as the function of frequency is plotted in Fig.~\ref{g-PForce} for different driving amplitudes $F$. Note that the power $P$ is normalized to the squared driving amplitude, $F^2$. For all the studied driving amplitudes the normalized power in the acoustic phonon band is the same. On the other hand, in the optic band the normalized power does not depend on $F$ for $F<0.08$~nN, but for higher values of $F$ power reduces with growing $F$. The inset in Fig.~\ref{g-PForce} shows the optic phonon band in an enlarged scale. At the same time, for the amplitudes $F>0.16$~nN the region of nonzero power appears inside the phonon gap close to the upper edge of the gap, which is at 35.37~THz.

For small frequencies ($\nu<5$~THz) power is frequency-independent in the case of force driving, see Fig.~\ref{g-PForce}. This is because the increase in driving frequency by a factor of $k$ results in the emission of phonon wave with the amplitude reduced by the factor of $k$, so that the product $A^2\nu^2$ remains unchanged, and this product is proportional to the phonon energy density.

\begin{figure}
\includegraphics*[scale=0.35]{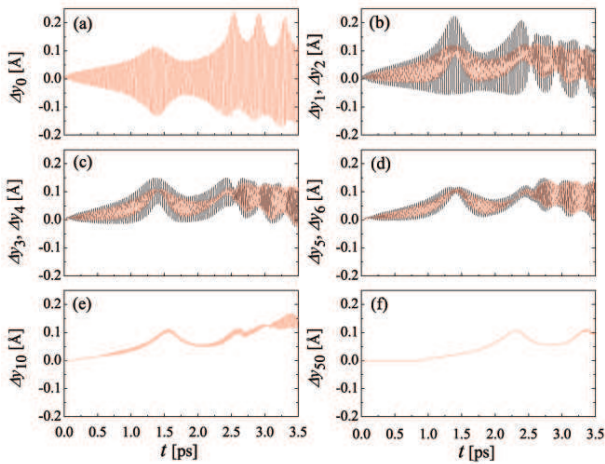}
\caption{\label{g-osc3}(Color online) Time dependence of $\Delta y_n$ for the atomic rows (a) $n=0$ (driven atoms), (b) $n=1$ and 2, (c) $n=3$ and 4, (d) $n=5$ and 6, (e) $n=10$, and (f) $n=50$. Curves for odd (even) atoms are shown in black (red). Force driving amplitude is $F=0.32~nN$ and frequency is $\nu=35.07$~THz.}
\end{figure}

In order to reveal the mechanism of energy transfer in the gap frequency region, in Fig.~\ref{g-osc3} we plot $\Delta y_n(t)$, $n=\{0,1,2,3,4,5,6,10,50\}$ for the driving amplitude $F=0.32~nN$ and frequency $\nu=35.07$~THz, which is inside the gap. The amplitude of the driven atom in (a) gradually increases with time and then the beating regime is extablished with the amplitude varying between 0.1 and 0.2 \AA. A remarkable difference between force driving and displacement driving is that in the former case the vibrations with the frequencies close to the driving frequency propagate deeper along the nanoribbon. Indeed, in Fig.~\ref{g-osc1}(d) and Fig.~\ref{g-osc2}(d) the row with $n=6$ (red curve) vibrates with low frequency, while in Fig.~\ref{g-osc3}(d) it vibrates at high frequency close to the driving frequency, which is also the frequency of gap DBs. This should be interpreted as the propagation of DBs along the nanoribbon in the case of force driving, which was not observed for displacement driving. At the same time, for the remote row with $n=50$ in Fig.~\ref{g-osc3}~(f), only the low-frequency oscillations can be seen. These oscillations appear due to the modulation of the amplitude of the driven atomic row. The long waves propagate much faster than DBs, that is why the high-frequency vibrations cannot be seen for the row $n=50$ by the end of the simulations run, $t=3.5$~ps.

We conclude that in the case of force driving with the inside gap frequency, the energy flow along the nanoribbon is due to propagation of both DBs and the low-frequency waves emitted due to the time modulation of DB amplitude. In the case of displacement driving only low-frequency waves were the energy carriers.

\section{Discussion and conclusions}
\label{Conclusion}
Molecular dynamics simulations of the energy transport in the strained graphene nanoribbon away from the ac driven atomic rows have been performed in quasi-one-dimensional setting. Driven zigzag atomic rows were considered as the energy source and the power of the energy source was calculated as the function of driving amplitude and frequency. The cases of displacement driving and force driving are compared and contrasted.

Our main results can be formulated as follows.

1. When driving frequency is {\em within the phonon bands}, there exist critical value of the driving amplitude below which the power normalized to the squared driving amplitude does not depend on the amplitude. For the displacement driving this is true for $A<0.01$~\AA$\,$ (see Fig.~\ref{g-PDisp2}), while for the force driving this holds for $F<0.08$~nN (see Fig.~\ref{g-PForce}). It is natural to call driving at such small amplitudes as {\em linear driving regime}. For larger driving amplitudes, due to anharmonicity of the interatomic bonds, energy density of the waves emitted by the driven atoms is no longer proportional to squared amplitude (to $A^2$ and $F^2$ for the displacement and force driving, respectively). Also note that for the acoustic waves the linear regime is observed for higher values of the driving amplitudes than for optic waves.

2. For driving frequencies {\em within the phonon gap}, in the case of displacement driving, even for the {\em linear driving regime} there exists a range of frequencies with nonzero power (see Fig.~\ref{g-PDisp2}). This is a nontrivial observation because typically the supratransmission at frequencies outside the phonon spectrum is observed in the nonlinear regime, i.e., for sufficiently large driving amplitudes \cite{GL2002,KLR2004,KTDBA2004}. The mechanism of the energy transport in the linear regime of displacement driving is related to the excitation of two standing DBs to the both sides of the driven zigzag atomic row. The amplitude of these DBs is time modulated and due to the effect of local "thermal expansion" such DBs emit low-frequency waves with the frequency of DB modulation (see Fig.~\ref{g-osc1} and Fig.~\ref{g-osc2}).

3. For driving frequencies {\em within the phonon gap}, in the case of force driving, the range of frequencies with nonzero power appears close to the upper edge of the phonon gap only in the nonlinear driving regime (see Fig.~\ref{g-PForce}). In this case the energy flow along the nanoribbon is due to propagation of both DBs and the low-frequency waves emitted due to the time modulation of DB amplitude.

4. No energy transport was observed for driving frequencies {\em above the phonon spectrum}. This is true for both displacement driving and force driving, see Fig.~\ref{g-PDisp2} and Fig.~\ref{g-PForce}, respectively. The explanation is that strained graphene does not support breathers with frequencies {\em above the phonon spectrum} \cite{myAbGrE2016}. Instead of that, gap DBs (with frequencies within the gap of phonon spectrum) do exist in graphene under the strain \cite{KhaDmiKiv2011} and they assist energy transport at driving frequencies within the gap.

The present study can be continued in many ways. Bearing in mind the discovery of transverse DBs in unstrained graphene \cite{DBnormal} it would be interesting to apply out-of-plane ac external driving to see if such DBs can manifest themselves in the energy transport. It is also timely to analyse the response of other two-dimensional and three-dimensional crystals to ac external driving in order to see the role of DBs in the energy absorption. Such studies would eventually suggest a clear experimental setup for indirect observation of DBs in crystals.\\

\begin{acknowledgments}
This work was supported by the Russian Science Foundation (Project 14-29-00199).
\end{acknowledgments}

\bibliography{references}{}
\end{document}